# Is the acceleration scale of the radial acceleration relation tracking the cosmic expansion rate?

G. T. Alcock ★

*Independent Researcher, Los Angeles, CA 90049, USA*

## ABSTRACT

The MUSE–DARK survey has reported a highly significant increase of the radial acceleration relation (RAR) scale $a_0$ with redshift over $0.33 < z < 1.44$, parametrized linearly and described by its authors as phenomenological. We confront the same binned measurements with physically motivated scalings, using the continuous family $a_0(z) = A\,(1+z)^\gamma$ as the primary statistic. We find $\gamma = 0.78 \pm 0.15$ (statistical, plotted uncertainties read as $1\sigma$): a redshift-independent scale ($\gamma = 0$) is excluded at $5.3\sigma$ by a nested test, and the matter–density scaling ($\gamma = 3/2$) at $4.8\sigma$. Hubble tracking, $a_0(z) \propto H(z)$, with effective exponent $\gamma \simeq 1.10$, is consistent with the measurement within $2.1\sigma$ and is the information-criterion-preferred one-parameter description; its amplitude, $A = (1.40 \pm 0.03) \times 10^{-10}\,\mathrm{m\,s^{-2}}$, lies 17 per cent above the canonical SPARC value, within the latter's systematic-dominated uncertainty. The survey's remark that the evolution is 'faster than $H(z)$' is shown to be anchor-dependent: relative to a floating amplitude the binned growth is, if anything, mildly *shallower* than $H(z)$. All exponent-based conclusions are invariant under redshift-independent rescalings of $a_0$, and hence robust to common-mode stellar-mass systematics. At current precision, the long-noted coincidence $a_0 \sim cH_0$ survives its first confrontation with direct kinematic measurements at intermediate redshift.

## 1 INTRODUCTION

The radial acceleration relation (RAR) is a tight empirical correlation between the total centripetal acceleration $g_{\mathrm{obs}}$ and the acceleration $g_{\mathrm{bar}}$ sourced by baryons alone (S. S. McGaugh, F. Lelli & J. M. Schombert 2016; F. Lelli et al. 2017). Its characteristic scale, $a_0 \simeq 1.2 \times 10^{-10}\,\mathrm{m\,s^{-2}}$, marks the acceleration below which observed dynamics depart from the baryonic Newtonian expectation. Whether $a_0$ is a fixed constant of nature, an emergent scale of galaxy formation, or a dynamical quantity tied to the cosmological background is a discriminating question for the missing-mass problem (B. Famaey & S. S. McGaugh 2012).

A numerical coincidence has been noted since the earliest days of modified Newtonian dynamics (MOND): $a_0$ is of the order of the cosmic acceleration scale, $a_0 \sim cH_0$ (M. Milgrom 1983), refined in later work to $a_0 \approx cH_0/2\pi$ (B. Famaey & S. S. McGaugh 2012; M. Milgrom 2020), with the possibility that the agreement reflects a vacuum or cosmological origin of the scale (M. Milgrom 1999) – a paper that already raised the possibility that $a_0$ varies with cosmic time (see also M. Milgrom 1983). If the proximity is causal rather than accidental, the most economical expectation is that the RAR scale tracks the expansion rate,

$$a_0(z) \;=\; A\,E(z), \qquad E(z) \equiv H(z)/H_0, \tag{1}$$

with $A$ a constant amplitude; the alternative reading, in which $a_0$ is tied to the cosmological constant (M. Milgrom 1999), would instead imply an essentially frozen scale at late times (M. Milgrom 2020). Because $E(z)$ increases monotonically with redshift, equation (1) predicts that the RAR scale was *larger* in the past. An early test against the redshift evolution of the Tully–Fisher relation found no supporting evidence within the systematics of the day (C. Limbach, D. Psaltis & F. Özel 2008), and redshift dependence of the RAR has since been discussed as a theoretical discriminant by several authors (S. Hossenfelder & T. Mistele 2018; V. G. Gueorguiev 2024). At low redshift, MIGHTEE-H I kinematics show a mild ($2.4\sigma$) preference for an increasing $a_0$ (A.-A. Vărăşteanu et al. 2025).

Direct kinematic measurements have now reached the regime where the question becomes quantitative. B. I. Ciocan et al. (2026b), using disc–halo decompositions of 79 star-forming galaxies at $0.33 < z < 1.44$ from the MUSE–DARK survey (B. I. Ciocan et al. 2026a), report that the RAR persists at intermediate redshift with a characteristic scale that increases systematically with $z$: adopting the linear parametrization $a_0(z) = a_0(0) + a_1 z$, they measure $a_1 = 1.59^{+0.11}_{-0.10} \times 10^{-10}\,\mathrm{m\,s^{-2}}$ (95 per cent confidence). The authors emphasize that the linear form is 'intended as a simple phenomenological description... rather than to provide a physically motivated description', and their comparison with physical scalings of $a_0$ is limited to the remark that the measured evolution 'is faster than that of $H(z)$'; their conclusions call for efforts to 'refine predictive models and explore the physical

★ E-mail: gary@gtacompanies.com

**Table 1.** Binned RAR scale from B. I. Ciocan et al. (2026b, fig. 3), digitized as described in the text, in units of $10^{-10}$ m s$^{-2}$.

| $z$ (plotted) | $a_0$ | Lower bar | Upper bar |
|---|---|---|---|
| 0.503 | 1.990 | 0.085 | 0.086 |
| 0.825 | 2.200 | 0.105 | 0.105 |
| 1.047 | 2.571 | 0.114 | 0.110 |
| 1.276 | 2.710 | 0.134 | 0.138 |

mechanisms that could shape any evolution of the RAR with cosmic time'.

This letter answers that call at the level the binned measurements currently permit, confronting them with the constant-$a_0$ hypothesis, Hubble tracking (equation 1), the matter–density scaling $a_0 \propto \sqrt{\bar{\rho}_m} \propto (1+z)^{3/2}$, and the continuous exponent family that contains all three as special cases.

## 2 DATA

We use the binned RAR-scale measurements of B. I. Ciocan et al. (2026b, their fig. 3): four quantile bins in redshift, fitted by those authors with the same interpolation-function machinery as their global analysis. The two extreme values (1.99 and $2.71 \times 10^{-10}$ m s$^{-2}$) are printed in their text; the intermediate values and all uncertainties are not tabulated and were digitized from their fig. 3 with axis-calibrated precision of $\sim 0.01 \times 10^{-10}$ m s$^{-2}$ (the two printed values are recovered exactly; the digitization uncertainty is $\lesssim$10 per cent of each error bar and is neglected). Table 1 lists the adopted points with their asymmetric bar endpoints, so that every input to the fits below is auditable.

The confidence level of the plotted bars is not stated in the source; since the quantitative fits of B. I. Ciocan et al. (2026b) are quoted at 95 per cent confidence, we carry out all fits under both readings – bars as $1\sigma$ and bars as 95 per cent intervals ($\sigma$ = bar/1.96) – and quote below which statements are and are not sensitive to the choice. The four bins are disjoint subsets of the sample, but share the parent analysis's stellar-mass scale, mass-to-light, inclination, and distance calibrations; our fits treat them as independent, an assumption we return to in Section 4. As the local anchor, we adopt the SPARC value $a_0 = 1.20 \times 10^{-10}$ m s$^{-2}$ with statistical uncertainty $\pm 0.02$ and systematic uncertainty $\pm 0.24$ (S. S. McGaugh et al. 2016), the reference used by B. I. Ciocan et al. (2026b); fits are reported with and without this $z \simeq 0$ datum (combined uncertainty 0.26; systematic-dominated). Following B. I. Ciocan et al. (2026a) we adopt a flat Lambda cold dark matter ($\Lambda$CDM) background with $\Omega_m = 0.307$ and $H_0 = 67.7$ km s$^{-1}$ Mpc$^{-1}$ (Planck Collaboration XIII, 2016); varying ($\Omega_m$, $H_0$) within current constraints changes $E(z)$ by <1 per cent over $0 < z < 1.5$ and none of what follows.

Our fits are performed on the binned relation; the global fit of B. I. Ciocan et al. (2026b) used the individual resolved points, and a per-galaxy re-analysis with physically motivated scalings imposed – possible with the survey's public data products – would supersede the binned comparison presented here. Within each bin $E(z)$ is nearly linear (bin widths $\Delta z \simeq 0.2$–0.3), so evaluating models at the plotted abscissae introduces only second-order bias in the shape comparison.

**Table 2.** Fits to the binned $a_0(z)$, with plotted bars read as $1\sigma$ ('c68') and as 95 per cent intervals ('c95'; every $\chi^2$ scales by $1.96^2$). Amplitudes in $10^{-10}$ m s$^{-2}$. AICc is quoted for the c68 reading; $\Delta$AICc is relative to the best model.

| Model | $k$ | $\chi^2/\nu$ (c68) | AICc | Parameters |
|---|---|---|---|---|
| | | *Four bins* | | |
| $a_0 = A$ (frozen) | 1 | 28.8/3 | 32.8 | $A = 2.28$ |
| $a_0 = A\,E(z)$ | 1 | 5.2/3 | 9.2 | $A = 1.40 \pm 0.03$ |
| $a_0 = A\,(1+z)^{3/2}$ | 1 | 23.2/3 | 27.2 | $A = 0.90 \pm 0.02$ |
| $a_0 = a + bz$ | 2 | 1.1/2 | 17.1 | $a = 1.48$, $b = 0.98$ |
| $a_0 = A\,(1+z)^{\gamma}$ | 2 | 1.2/2 | 17.2 | $\gamma = 0.78 \pm 0.15$ |
| | | *Four bins + SPARC anchor* | | |
| $a_0 = A$ (frozen) | 1 | 45.4/4 | 48.7 | $A = 2.24$ |
| $a_0 = A\,E(z)$ | 1 | 5.8/4 | 9.1 | $A = 1.40 \pm 0.03$ |
| $a_0 = A\,(1+z)^{3/2}$ | 1 | 24.6/4 | 27.9 | $A = 0.90 \pm 0.02$ |
| $a_0 = A\,(1+z)^{\gamma}$ | 2 | 1.8/3 | 11.8 | $\gamma = 0.83 \pm 0.13$ |

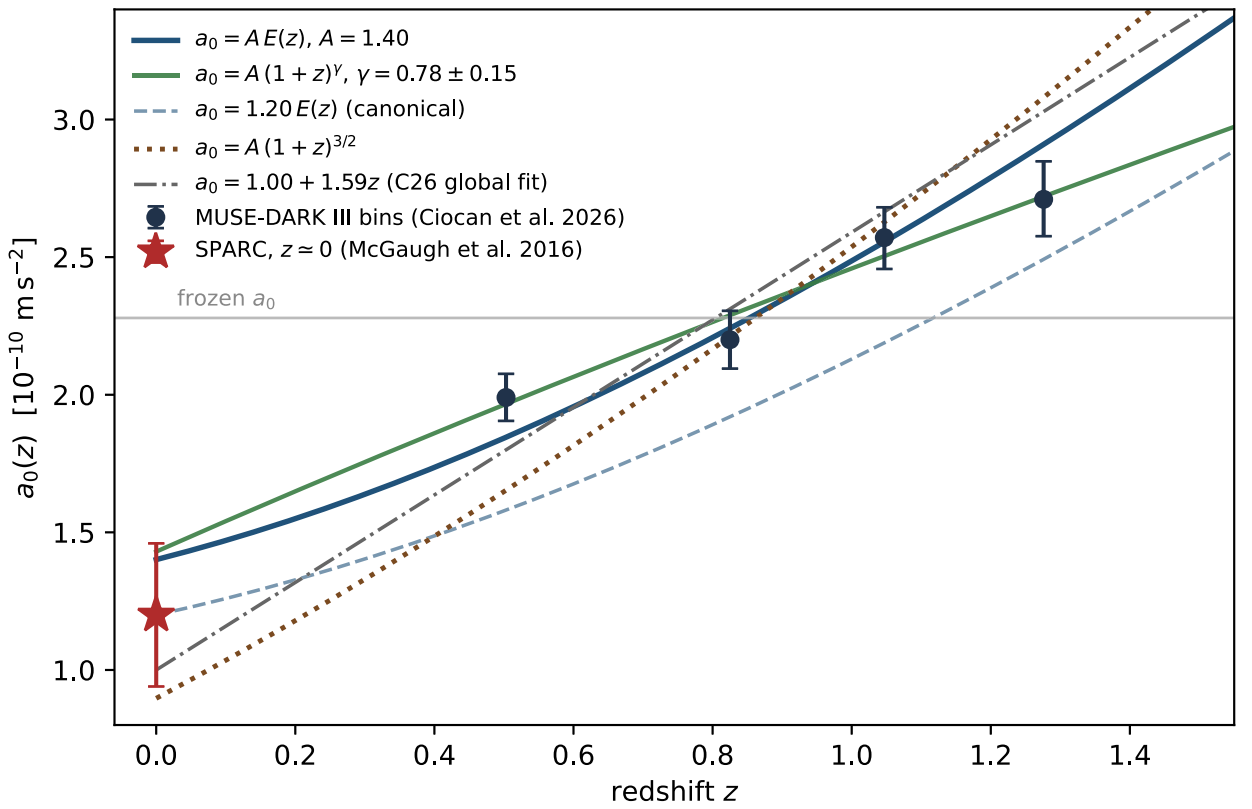

**Figure 1.** The binned RAR acceleration scale of B. I. Ciocan et al. (2026b; circles; table 1), the local SPARC value (star; S. S. McGaugh et al. 2016), and the models of Table 2. Hubble tracking with a free amplitude (solid blue) and the best-fitting exponent family (solid green, $\gamma = 0.78$) describe the binned evolution; the canonical-amplitude curve (dashed), the matter–density scaling (dotted) and a frozen $a_0$ (horizontal line) do not.

## 3 MODELS AND RESULTS

We fit four hypotheses by weighted least squares: a redshift-independent scale ($a_0 = A$); Hubble tracking (equation 1); the matter–density scaling $a_0 = A\,(1+z)^{3/2}$; and the linear form of B. I. Ciocan et al. (2026b). As the primary statistic we fit the continuous exponent family

$$a_0(z) \;=\; A\,(1+z)^{\gamma}, \tag{2}$$

which contains the frozen ($\gamma = 0$) and matter–density ($\gamma = 3/2$) hypotheses as special cases and closely brackets Hubble tracking: over $0.5 \lesssim z \lesssim 1.3$ the effective logarithmic slope of $E(z)$ is $\gamma_E \simeq 1.10$. Model selection uses the small-sample-corrected Akaike criterion AICc $= \chi^2 + 2k + 2k(k+1)/(N-k-1)$, appropriate at $N$ = 4–5 points; significances for nested hypotheses use $\Delta\chi^2$ with one degree of freedom, and absolute goodness of fit (GOF) is quoted where relevant. Table 2 collects the results and Fig. 1 shows the data and models. All significances quoted in this section are statistical only (Section 4).

Four results follow:

**(i) Evolution is required:** The frozen model gives an absolute GOF of $\chi^2 = 28.8$ for three degrees of freedom ($p = 2.4 \times 10^{-6}$, $4.6\sigma$) against the bins alone. The properly nested test – $b = 0$ within the linear family – gives $\Delta\chi^2 = 27.7$ for one degree of freedom and a $5.3\sigma$ rejection (rising to $6.6\sigma$ with the anchor, and to $\gtrsim 10\sigma$ under the c95 reading). Equivalently, the fitted exponent $\gamma = 0.78 \pm 0.15$ excludes $\gamma = 0$ at $5.3\sigma$. This confirms, in functional-form language, the evolution detected by B. I. Ciocan et al. (2026b), and directly disfavours interpretations in which $a_0$ is a fixed constant of nature – including readings that tie $a_0$ to the cosmological constant and hence predict a frozen late-time scale.

**(ii) The matter–density scaling is excluded:** The fitted exponent excludes $\gamma = 3/2$ at $4.8\sigma$ (c68; $9.5\sigma$ under c95): the measured evolution is significantly shallower than $\sqrt{\bar{\rho}_{\rm m}}$-tracking.

**(iii) Hubble tracking is the preferred one-parameter description, with a convention caveat:** Under the c68 reading, $a_0 = A\,E(z)$ gives $\chi^2 = 5.2/3$ (GOF, $p = 0.16$) with no bin deviating by more than $1.7\sigma$, and is decisively preferred by AICc (9.2, against 17.1–17.2 for the two-parameter forms and 27–33 for the alternatives); its amplitude is $A = (1.40 \pm 0.03_{\rm stat}) \times 10^{-10}\,{\rm m\,s^{-2}} \simeq 0.21\,cH_0$, about a third above the $cH_0/2\pi$ coincidence value (B. Famaey & S. S. McGaugh 2012), and 17 per cent above the canonical SPARC amplitude – a $0.8\sigma$ difference against SPARC's systematic-dominated uncertainty, though we caution that treating that systematic as Gaussian is a convention. The fitted exponent sits $2.1\sigma$ below $\gamma_E \simeq 1.10$. Under the c95 reading the same central values hold but the statistical bars shrink, pure $E(z)$ tracking is then in GOF tension ($\chi^2 = 19.9/3$, $p = 1.8 \times 10^{-4}$) and the exponent family prefers $\gamma = 0.78 \pm 0.08$, i.e. growth shallower than $E(z)$ at $\sim 4\sigma$. Resolving the plotted confidence convention – or, better, refitting the individual galaxies – therefore determines whether the data *select* $E(z)$-tracking or merely *bracket* it; the exclusions of $\gamma = 0$ and $\gamma = 3/2$ hold under either reading.

**(iv) 'Faster than $H(z)$' is anchor-dependent and the binned relation, if anything, favours the opposite:** Relative to the canonical local value $1.20 \times 10^{-10}\,{\rm m\,s^{-2}}$ taken as exact, the growth to $z \simeq 1$ indeed exceeds $E(z)$, consistent with the remark of B. I. Ciocan et al. (2026b); the same is true relative to their fitted intercept $a_0(0) = 1.00 \times 10^{-10}\,{\rm m\,s^{-2}}$. But the exponent of the binned relation itself, with the amplitude free, is $\gamma = 0.78 \pm 0.15$ – *shallower* than $E(z)$'s effective exponent at $2.1\sigma$. The 'faster than $H(z)$' characterization is thus a statement about the intercept, not about the measured growth rate: once the local anchor is allowed its published systematic freedom, Hubble tracking is not excluded, and the residual pull of the binned shape is mildly toward slower, not faster, growth.

## 4 ROBUSTNESS

The dominant systematic identified by B. I. Ciocan et al. (2026b) is the stellar-mass scale: reconciling all bins with the canonical local $a_0$ would require masses larger by $+0.2$ to $+0.45$ dex, shifts they argue are unsupported by independent checks. In the low-acceleration regime the fitted scale responds inversely to the mass normalization, so a common-mode mass offset rescales all bins coherently by tens of per cent. We therefore separate our conclusions by their exposure. *Exponent-based conclusions – the exclusions of $\gamma = 0$ and $\gamma = 3/2$, and the measured $\gamma$ itself – are exactly invariant under any redshift-independent multiplicative rescaling of $a_0$*, because such a rescaling is absorbed entirely by the free amplitude. They are exposed only to redshift-*dependent* systematics (e.g. stellar-mass estimates degrading with $z$), which the survey's appendices bound but do not eliminate. *Amplitude-based statements* – the value of $A$, its 17 per cent offset from SPARC, and the rejection of the fixed canonical amplitude – inherit the full common-mode uncertainty and should be read as conditional on the published mass scale. The bins also share calibration machinery, so our independence assumption is optimistic for the absolute significances; the tabulated bin values, their confidence convention, and any covariance estimate from the survey team would sharpen every number in Table 2.

A further caveat specific to modified-dynamics readings is the external field effect (EFE): per-galaxy RAR fits require an environment-correlated external-field term, so the effective $a_0$ inferred galaxy-by-galaxy depends on the ambient field (K.-H. Chae et al. 2020), and both the mean cosmic field and typical environmental fields evolve with redshift, so an evolving EFE can in principle mimic an evolving apparent $a_0$. B. I. Ciocan et al. (2026b) exclude interacting galaxies and cluster members and argue the EFE impact is minor for their sample; we note only that the sign of the effect – weaker external fields, hence weaker apparent-$a_0$ suppression, at higher $z$ – would act to *steepen* the apparent evolution, in the direction of the measured signal, and deserves quantitative treatment in a per-galaxy analysis.

## 5 DISCUSSION

The MUSE–DARK III measurement is the first direct kinematic detection of an evolving RAR scale at intermediate redshift, and the analysis above shows that its functional form is consistent, at current precision, with the simplest dynamical reading of the oldest numerical coincidence in the field: $a_0(z) \sim cH(z)$ (M. Milgrom 1983, 1999). Three implications follow:

First, for modified-dynamics interpretations, a strictly constant $a_0$ is now under direct observational pressure at the $\gtrsim 5\sigma$ (statistical) level, subject to the redshift-dependent systematics discussed above; frameworks tying $a_0$ to the cosmological constant inherit the same pressure, while mechanisms tying the transition scale to the evolving background are qualitatively favoured (M. Milgrom 1999; S. Hossenfelder & T. Mistele 2018). The early Tully–Fisher test of C. Limbach et al. (2008), which disfavoured a naive $cH(z)$ scaling within the systematics then available, is superseded in this regime by direct kinematics; the scale-invariant-vacuum analysis of V. G. Gueorguiev (2024), which found a $z$-slope of $a_0$ consistent with zero in earlier data, is likewise in tension with the MUSE–DARK detection, as those authors' framework predicts slower evolution than is now measured. We also note that the apparently opposing literature on declining high-$z$ rotation curves and baryon-dominated discs (R. Genzel et al. 2017; P. Lang et al. 2017) – which was read as a *weaker* missing-mass effect at high $z$, with a MOND response given by M. Milgrom (2017) – probes larger radii and higher masses than the MUSE–DARK regime; reconciling the two data sets within any single $a_0(z)$ is an open task.

Secondly, for $\Lambda$CDM-based interpretations, an apparent $a_0(z)$ rising with redshift can plausibly emerge from the interplay of baryonic compactness, halo response, and feedback, as discussed by B. I. Ciocan et al. (2026b). The value of the present analysis is discriminative rather than interpretive: whatever produces the evolution must, over $0.3 \lesssim z \lesssim 1.4$, produce an effective exponent $\gamma \simeq 0.8 \pm 0.15$ – close to, and marginally shal-lower than, $E(z)$-tracking – a nontrivial quantitative target for simulation-based explanations.

Thirdly, the viable functional forms diverge rapidly beyond the measured range: at $z = 2$, Hubble tracking with the fitted amplitude predicts $a_0 \simeq 3.2 \times 10^{-10}\,\mathrm{m\,s^{-2}}$ while the linear form predicts 4.2; at $z = 3$, 4.1 versus 5.8. *James Webb Space Telescope* (*JWST)* and Atacama Large Millimeter/submillimeter Array (ALMA) rotation curves at $z \gtrsim 2$, and ultimately SKA H I kinematics, can separate them. A per-galaxy re-analysis of the MUSE–DARK sample with $E(z)$-tracking imposed in the global fit – which the survey's public data products make possible – would sharpen every comparison made here, and we recommend it to the survey team, together with publication of the tabulated bin values and their confidence convention.

The coincidence $a_0 \sim cH_0$ has stood for four decades as a curiosity. The first direct kinematic measurements of the RAR scale across cosmic time neither break it nor confirm it: they are consistent with $a_0$ tracking $H(z)$ at current precision, and they already exclude the two simplest alternatives on either side.

## ACKNOWLEDGEMENTS

This work uses published measurements from the MUSE–DARK survey (B. I. Ciocan et al. 2026a, b) and the SPARC results of S. S. McGaugh et al. (2016). No new observational data were taken.

## DATA AVAILABILITY

All inputs are tabulated in Table 1; the fitting scripts and the digitization record are provided as supplementary material accompanying the submission.

## SUPPLEMENTARY MATERIAL

Supplementary material are available at *MNRAS* online.

**supplementary_material**

Please note: Oxford University Press is not responsible for the content or functionality of any supporting materials supplied by the authors. Any queries (other than missing material) should be directed to the corresponding author for the article.